\documentclass[aps,prl,preprint,tightenlines,superscriptaddress,showpacs,byrevtex]{revtex4}

\def \ks {K_{\rm S}^0}
\usepackage{graphicx}
\usepackage{color}

\begin{document}

\preprint{\vbox{
\hbox{Belle Preprint 2006-16}
\hbox{KEK   Preprint 2006-12}
}}

\title{ \quad\\[0.5cm] Search for Lepton Flavor Violating $\tau^-$ Decays
{with a} $\ks$ Meson}

\affiliation{Budker Institute of Nuclear Physics, Novosibirsk, Russia}
\affiliation{Chiba University, Chiba, Japan}
\affiliation{University of Cincinnati, Cincinnati, OH, USA}
\affiliation{University of Hawaii, Honolulu, HI, USA}
\affiliation{High Energy Accelerator Research Organization (KEK), Tsukuba, Japan}
\affiliation{University of Illinois at Urbana-Champaign, Urbana, IL, USA}
\affiliation{Institute for High Energy Physics, Protvino, Russia}
\affiliation{Institute of High Energy Physics, Vienna, Austria}
\affiliation{Institute for Theoretical and Experimental Physics, Moscow, 
Russia}
\affiliation{J. Stefan Institute, Ljubljana, Slovenia}
\affiliation{Kanagawa University, Yokohama, Japan}
\affiliation{Korea University, Seoul, South Korea}
\affiliation{Kyungpook National University, Taegu, South Korea}
\affiliation{Swiss Federal Institute of Technology of Lausanne, EPFL, Lausanne, Switzerland}
\affiliation{University of Maribor, Maribor, Slovenia}
\affiliation{University of Melbourne, Victoria, Australia}
\affiliation{Nagoya University, Nagoya, Japan}
\affiliation{Nara Women's University, Nara, Japan}
\affiliation{National Central University, Chung-li, Taiwan}
\affiliation{National United University, Miao Li, Taiwan}
\affiliation{Department of Physics, National Taiwan University, Taipei, 
Taiwan}
\affiliation{H. Niewodniczanski Institute of Nuclear Physics, Krakow, Poland}
\affiliation{Nippon Dental University, Niigata, Japan}
\affiliation{Niigata University, Niigata, Japan}
\affiliation{Nova Gorica Polytechnic, Nova Gorica, Slovenia}
\affiliation{Osaka City University, Osaka, Japan}
\affiliation{Osaka University, Osaka, Japan}
\affiliation{Panjab University, Chandigarh, India}
\affiliation{Peking University, Beijing, PR China}
\affiliation{Princeton University, Princeton, NJ, USA}
\affiliation{RIKEN BNL Research Center, Brookhaven, NY, USA}
\affiliation{University of Science and Technology of China, Hefei, PR China}
\affiliation{Seoul National University, Seoul, South Korea}
\affiliation{Sungkyunkwan University, Suwon, South Korea}
\affiliation{University of Sydney, Sydney, NSW, Australia}
\affiliation{Toho University, Funabashi, Japan}
\affiliation{Tohoku Gakuin University, Tagajo, Japan}
\affiliation{Tohoku University, Sendai, Japan}
\affiliation{Department of Physics, University of Tokyo, Tokyo, Japan}
\affiliation{Tokyo Institute of Technology, Tokyo, Japan}
\affiliation{Tokyo Metropolitan University, Tokyo, Japan}
\affiliation{Tokyo University of Agriculture and Technology, Tokyo, Japan}
\affiliation{Virginia Polytechnic Institute and State University, Blacksburg, VA, USA}
\affiliation{Yonsei University, Seoul, South Korea}

 \author{Y.~Miyazaki}\affiliation{Nagoya University, Nagoya, Japan}  
 \author{K.~Abe}\affiliation{High Energy Accelerator Research Organization (KEK), Tsukuba, Japan}  
 \author{K.~Abe}\affiliation{Tohoku Gakuin University, Tagajo, Japan} 
 \author{H.~Aihara}\affiliation{Department of Physics, University of Tokyo, Tokyo, Japan} 
 \author{D.~Anipko}\affiliation{Budker Institute of Nuclear Physics,
 Novosibirsk, Russia} 
 \author{K.~Arinstein}\affiliation{Budker Institute of Nuclear Physics,
 Novosibirsk, Russia} 
 \author{V.~Aulchenko}\affiliation{Budker Institute of Nuclear Physics,
 Novosibirsk, Russia} 
 \author{T.~Aushev}\affiliation{Institute for Theoretical and
 Experimental Physics, Moscow, Russia}  
 \author{A.~M.~Bakich}\affiliation{University of Sydney, Sydney, NSW, Australia} 
 \author{M.~Barbero}\affiliation{University of Hawaii, Honolulu, HI, USA} 
% ...
\collaboration{The Belle Collaboration}

\begin{abstract}
We have searched for 
{the lepton flavor violating {decays}}
$\tau^-\rightarrow \ell^-\ks$ ($\ell = e \mbox{ or } \mu$),
using a data sample of
281 fb$^{-1}$ collected with
the Belle detector at the KEKB $e^+e^-$ asymmetric-energy collider.
No evidence for a signal {was} found
in either of the decay modes,
{and we} set the following upper limits 
for the branching {fractions:}
${\cal{B}}(\tau^-\rightarrow e^-\ks) < 5.6\times 10^{-8}$
and 
${\cal{B}}(\tau^-\rightarrow \mu^-\ks) < 4.9\times 10^{-8}$ 
at the 90\% confidence level. 
{These results 
improve the previously published limits
set by {the} CLEO collaboration 
by factors of 16 and 19, respectively.}
\end{abstract}


\maketitle

\section{Introduction}

{Lepton flavor violation (LFV) 
{is allowed} in many extensions of the Standard Model (SM),
{such as} Supersymmetry (SUSY) and leptoquark models.}
{In particular,} lepton flavor violating decays with $\ks$ mesons 
{are discussed} 
in models
{with heavy} singlet Dirac neutrinos~\cite{cite:amon}, 
$R-$parity violation in SUSY~\cite{cite:rpv, cite:rpv2},
dimension-six effective fermionic operators that induce $\tau-\mu$
mixing~\cite{cite:six_fremionic}.
Experiments at the  $B$-factories
allow 
searches
for {lepton flavor violating decays} with
a very high sensitivity.
The best upper limits {of}
${\cal{B}}(\tau^-\rightarrow e^-\ks) < 9.1\times 10^{-7}$
and 
${\cal{B}}(\tau^-\rightarrow \mu^-\ks) < 9.5\times 10^{-7}$
at the 90\% confidence level
{were set
by  the CLEO experiment}
using 13.9 fb${}^{-1}$ of data~\cite{cite:cleo}.

In this paper,
we {report}  a {search} for
{the lepton flavor violating  decays}
$\tau^-\rightarrow \ell^-\ks$
($\ell = e \mbox{ or } \mu$)\footnotemark[2]
{using  281 fb$^{-1}$ of data 
collected at the $\Upsilon(4S)$ resonance
and 60 MeV below it}
with the Belle detector at the KEKB  $e^+e^-$ 
asymmetric-energy collider~\cite{kekb}. 
\footnotetext[2]{Unless otherwise stated, charge 
conjugate decays are 
{included}
throughout
this paper.}

The Belle detector is a large-solid-angle magnetic spectrometer that
consists of a silicon vertex detector (SVD), 
a 50-layer central drift chamber (CDC), 
an array of aerogel threshold \v{C}erenkov counters (ACC), a barrel-like arrangement of 
time-of-flight scintillation counters (TOF), and an electromagnetic calorimeter 
{comprised of}  
CsI(Tl) {crystals (ECL), all located} inside
a superconducting solenoid coil
that provides a 1.5~T magnetic field.  
An iron flux-return located outside of the coil is instrumented to detect $K_{\rm{L}}^0$ mesons 
and to identify muons (KLM).  
The detector is described in detail elsewhere~\cite{Belle}.

{Particle identification 
is very important in this measurement.
{We use particle identification 
likelihood variables based on}}
the ratio of the energy 
deposited in the ECL to the momentum measured in the SVD and CDC, 
the shower shape in the ECL, 
the particle range in the KLM, 
the hit information from the ACC,
{the measured $dE/dX$} in the CDC 
and {the particle's time-of-flight} from the TOF.
{For lepton identification,
we {form}  
a likelihood ratio based on the}
electron probability ${\cal P}(e)$~\cite{EID} and 
{the} muon probability ${\cal P}({\mu})$~\cite{MUID} 
determined by
the responses of the appropriate subdetectors.

{For Monte Carlo (MC) {simulation} studies,}
the following programs have been used to
generate background events:
KORALB/TAUOLA~\cite{cite:koralb_tauola} for $\tau^+\tau^-$, 
QQ~\cite{cite:qq} for $B\bar{B}$ and continuum,
BHLUMI~\cite{BHLUMI} for {Bhabha events,}
KKMC~\cite{KKMC} for $e^+e^-\rightarrow\mu^+\mu^-$ and
AAFH~\cite{AAFH} for two-photon processes.
{Since the QQ generator does not include some rare processes 
{that potentially
contribute to 
final states with a $\ks$ meson,} 
we generated special samples of  
{$e^+e^- \to D^{*+}D^{(*)-}$, a process
that was}
recently observed by 
{the} 
Belle collaboration~\cite{uglov}.}
Signal MC is generated by KORALB/TAUOLA.
{Signal $\tau$ decays are  two-body 
and assumed}
to  have a uniform angular distribution
{in the $\tau$} lepton's rest frame. 
The Belle detector response is simulated by a GEANT 3~\cite{cite:geant3} 
based program.
{All kinematic variables are
calculated in the laboratory frame
{unless otherwise specified.}
In particular,
variables
calculated in the $e^+e^-$ center-of-mass (CM) frame 
are indicated by the superscript ``CM''.}

\section{Data Analysis}


{We search for $\tau^+\tau^-$ events
in which one $\tau$ (signal side) decays
into $\ell\ks$ ($\ks \to \pi^+\pi^-$),
while 
the other $\tau$ (tag side) decays 
into  one charged track 
(with a {sign} opposite to that of the
signal-side lepton)
and any number of additional photons and neutrinos.}
Thus, the experimental signature is:
\begin{center}
$\left\{
\tau^- \rightarrow \ell^-(=e^-\mbox{ or }\mu^-) + \ks(\rightarrow\pi^+\pi^-)
\right\} 
~+
~ \left\{ \tau^+ \rightarrow ({\rm a~track})^+ + (n^{\rm TAG} {\gamma} \ge 0)
 + X(\rm{missing}) 
\right\}$.
\end{center}
{All charged tracks} and photons 
are required to be reconstructed 
{within a fiducial volume,} 
defined by $-0.866 < \cos\theta < 0.956$,
where $\theta$ is the polar angle with
respect to the direction opposite to the $e^+$ beam.
{We select charged tracks with
{momenta} transverse to the $e^+$ beam
$p_t > 0.1$ GeV/$c$ 
and 
photons with energies
$E_{\gamma} > 0.1$ GeV.}

%
%
{Candidate $\tau$-pair events are required to have} 
four charged tracks {with}  a zero net charge.
{Events {are} separated into two 
hemispheres corresponding {to} 
the signal (three-prong)
and tag (one-prong) sides
by the plane perpendicular to the thrust
axis~\cite{thrust}. }
The
magnitude of the thrust is required to be larger than 0.9 to suppress
the $q\bar{q}$ continuum background. 
{The $\ks$ is 
reconstructed 
from two {oppositely-charged tracks in the signal side
that {have an invariant mass}   
{0.482 GeV/$c^2 < M_{\pi^+\pi^-} <0.514$ {GeV/$c^2$},
assuming a pion {mass} for both tracks.}}}
The $\pi^+\pi^-$ vertex is required to
be displaced from the interaction point (IP)
in the direction of the pion pair momentum~\cite{cite:ks}.
{In order to avoid fake $\ks$ candidates 
from  {photon} conversions
(i.e. $\gamma \rightarrow e^+e^-$),}
the invariant mass reconstructed 
by assigning the electron mass to the tracks,
is required to be greater than 0.2 GeV/$c^2$.
{The signal side track not used in the $\ks$ reconstruction
is required to satisfy 
the lepton identification selection.}
{The electron and muon {identification} criteria are}
${\cal P}(e) > 0.9$ with $p > 0.3$ GeV/$c$
and 
${\cal P}(\mu) > 0.9$ with $p > 0.6$ GeV/$c$,
respectively.
After the event selection described above, 
{most of the remaining background comes from
generic $\tau^+\tau^-$ and continuum events 
that contain a real $\ks$ meson.}

To ensure that the missing particles are neutrinos rather
than photons or charged particles 
{that fall outside the detector acceptance,} 
we impose additional requirements on the missing
momentum vector, $\vec{p}_{\rm miss}$, 
calculated by subtracting the
vector sum of the momenta
of
all tracks and photons 
from the sum of the $e^+$ and $e^-$ beam momenta.
We require that the magnitude of $\vec{p}_{\rm miss}$ 
{be} greater than
0.4 GeV/$c$ and that 
{its direction point into} 
the fiducial volume of the
detector,
as shown for {the} $\tau^-\rightarrow\mu^-\ks$ mode 
in Fig.~\ref{fig:cut} (a) and (b).
The total visible energy in the CM frame{,}  
{$E^{\mbox{\rm{\tiny{CM}}}}_{\rm{vis}}$,}
is defined as the sum of the energies
{of the $\ks$ candidate,
the lepton,
the tag-side track 
(with {a} pion mass hypothesis) 
and all photon candidates.
{We require $E^{\mbox{\rm{\tiny{CM}}}}_{\rm{vis}}$ 
to satisfy}
{the condition:}
$5.29$ GeV $< E^{\mbox{\rm{\tiny{CM}}}}_{\rm{vis}} < 10.0$ GeV 
(see Fig.~\ref{fig:cut} (c)).
{Since neutrinos are emitted only on the tag side,
the direction of 
{$\vec{p}_{\rm miss}$
should lie within the tag side of the event.}}
The cosine of the
opening angle between 
{$\vec{p}_{\rm miss}$}
and
{the} tag-side track 
in the CM system,
{$\cos \theta^{\mbox{\rm \tiny CM}}_{\rm tag-miss}$,}
{is therefore required to be greater than 0}
(see Fig.~\ref{fig:cut} (d)).
{For all kinematic distributions shown in Fig.~\ref{fig:cut},
{reasonable agreement between the data and background MC is observed.}}
{In order to suppress background 
from $q\bar{q}$ ($q = u, d, s, c$) 
continuum events,}
the following requirements 
on
the number of the photon candidates on the signal and tag side
{are imposed:}
{$n^{\rm{SIG}}\leq 1$ and $n^{\rm{TAG}}\leq 2$,}
respectively.

\begin{figure}[h]
\begin{center}
 \resizebox{0.7\textwidth}{0.6\textwidth}{\includegraphics
 {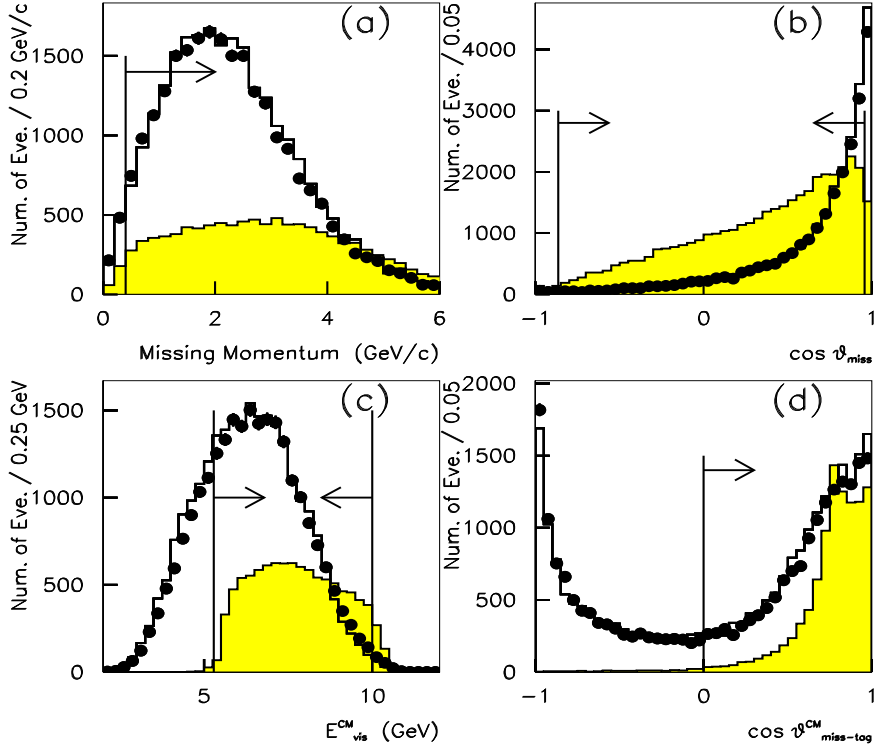}}
 \caption{ 
 Kinematic distributions used in the event selection
 after $\ks$ mass and muon {identification requirements:}
 (a) the momentum of the missing particle;
 (b) the polar angle of the missing particle; 
 (c) the total visible energy in the CM frame;
 (d) the opening angle between the missing particle and
 tag-side track in the CM frame.
 The signal MC distributions are indicated by the filled histograms, 
 {the total background} including 
 $\tau^+\tau^-$ and $q\bar{q}$ {is shown by} the open histogram, 
 and   
 {solid}
 circles are data.
 While the signal MC ($\tau^-\rightarrow\mu^-\ks$) 
 distribution is normalized arbitrarily, 
 {the data and background MC} are normalized to the same luminosity.
 {Selected regions are indicated  by arrows from the marked cut {boundaries.}} 
}
\label{fig:cut}
\end{center}
\end{figure}

Finally, 
the correlation between {the}
reconstructed momentum of {the} $\ell\ks$ system,
$p_{\ell K_{\rm S}}$,
and
the cosine of the opening angle
{between the lepton and $\ks$,}
$\cos \theta_{\ell K_{\rm S}}$
{is employed to further suppress
background from generic $\tau^+\tau^-$
and continuum events via the requirements:}
$\cos \theta_{\ell K_{\rm S}} < 0.14\times\log(p_{\ell K_{\rm S}}-2.7)+0.7$,
where $p_{\ell K_{\rm S}}$ is in GeV/$c$
(see Fig. \ref{fig:pmiss_vs_mmiss2}). 
While this condition retains
99\% of the signal,
99\% of the generic $\tau^+\tau^-$ and 84\% of {the} 
$uds$ continuum background
are removed.
{Following all the selection criteria,}
the signal detection efficiencies 
for {the} {$\tau^-\rightarrow e^-\ks$} and
{$\tau^-\rightarrow\mu^-\ks$} modes are
15.0\% and 16.2\%, respectively.

\begin{figure}[t]
\begin{center}
 \resizebox{.7\textwidth}{!}{\includegraphics
 {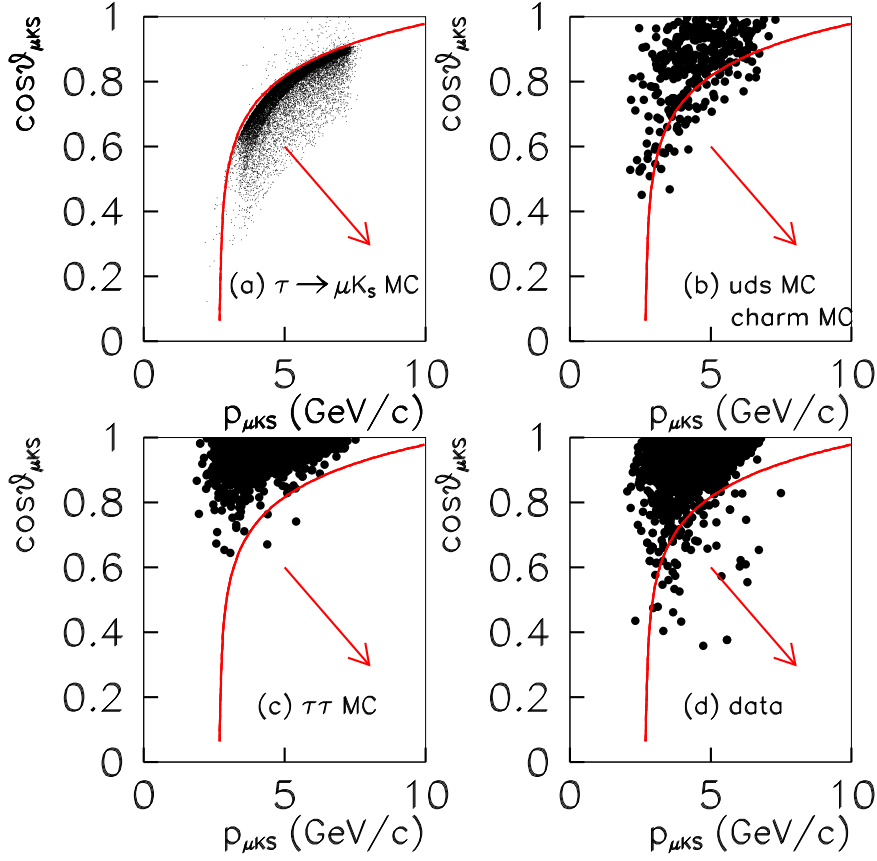}}
 \caption{
Scatter-plots 
 of (a) signal MC ($\tau^-\rightarrow\mu^-\ks$),  (b) continuum MC,  
(c) generic $\tau^+\tau^-$ MC events 
 and (d) data on {the}
$p_{\mu K_{\rm S}}$ vs $\cos \theta_{\mu K_{\rm S}}$ plane.
 Selected regions are indicated by curves with arrows.
 \label{fig:pmiss_vs_mmiss2}
 }
 \end{center}
\end{figure}

\section{Results}

Signal candidates are examined in the two-dimensional 
{plots} 
of the $\ell^-\ks$ invariant 
mass, $M_{{\ell\ks}}$, and the difference of their energy from the 
beam energy in the CM system, $\Delta E$.
A signal event should have $M_{{\ell\ks}}$
close to the $\tau$-lepton mass
and 
$\Delta E$ close to 0.
{For both modes,
the $M_{{\ell\ks}}$ and $\Delta E$  resolutions} are parameterized 
from the MC distributions around the peak  
{{with  bifurcated Gaussian shapes}
to account for initial state radiation.
These Gaussian have widths 
$\sigma^{\rm{high/ low}}_{M_{e\ks}} = 6.2/ 7.4$ MeV/$c$$^2$ and 
$\sigma^{\rm{high/ low}}_{\Delta E} = 20/ 26$ MeV
for the $\tau^-\rightarrow e^-\ks$ mode{,}
and
$\sigma^{\rm{high/ low}}_{M_{{\mu\ks}}} = 6.1/ 5.9$ MeV/$c$$^2$ and 
$\sigma^{\rm{high/ low}}_{\Delta E} = 19/ 23$ MeV
for the $\tau^-\rightarrow \mu^-\ks$ mode,}
where the ``high/low'' superscript indicates the higher/lower side 
of the peak.

We blind a region of $\pm 5\sigma_{{M_{\ell\ks}}}$ 
around the $\tau$ mass in $\rm{M_{\ell\ks}}$ 
and 
a region of
$-0.5\mbox{ GeV} < \Delta E < 0.5$ GeV
{so as not to bias our choice of selection criteria.}
Figure~\ref{fig:3} shows scatter-plots 
for data and signal MC samples 
distributed over $\pm 15\sigma$ 
in the $M_{{\ell\ks}}-\Delta E$ plane.
{Most of the surviving}
background events in both modes
come from 
$D^{\pm}\rightarrow\pi^{\pm}\ks$
and 
$D^{\pm}\rightarrow\ell^{\pm}\nu\ks$ decays.
The remaining continuum backgrounds in the $\tau^-\rightarrow \mu^-\ks$  mode
are combinations of a true $\ks$ meson and a fake lepton.

\begin{figure}[t]
\begin{center}
 \resizebox{0.496\textwidth}{0.496\textwidth}{\includegraphics
 {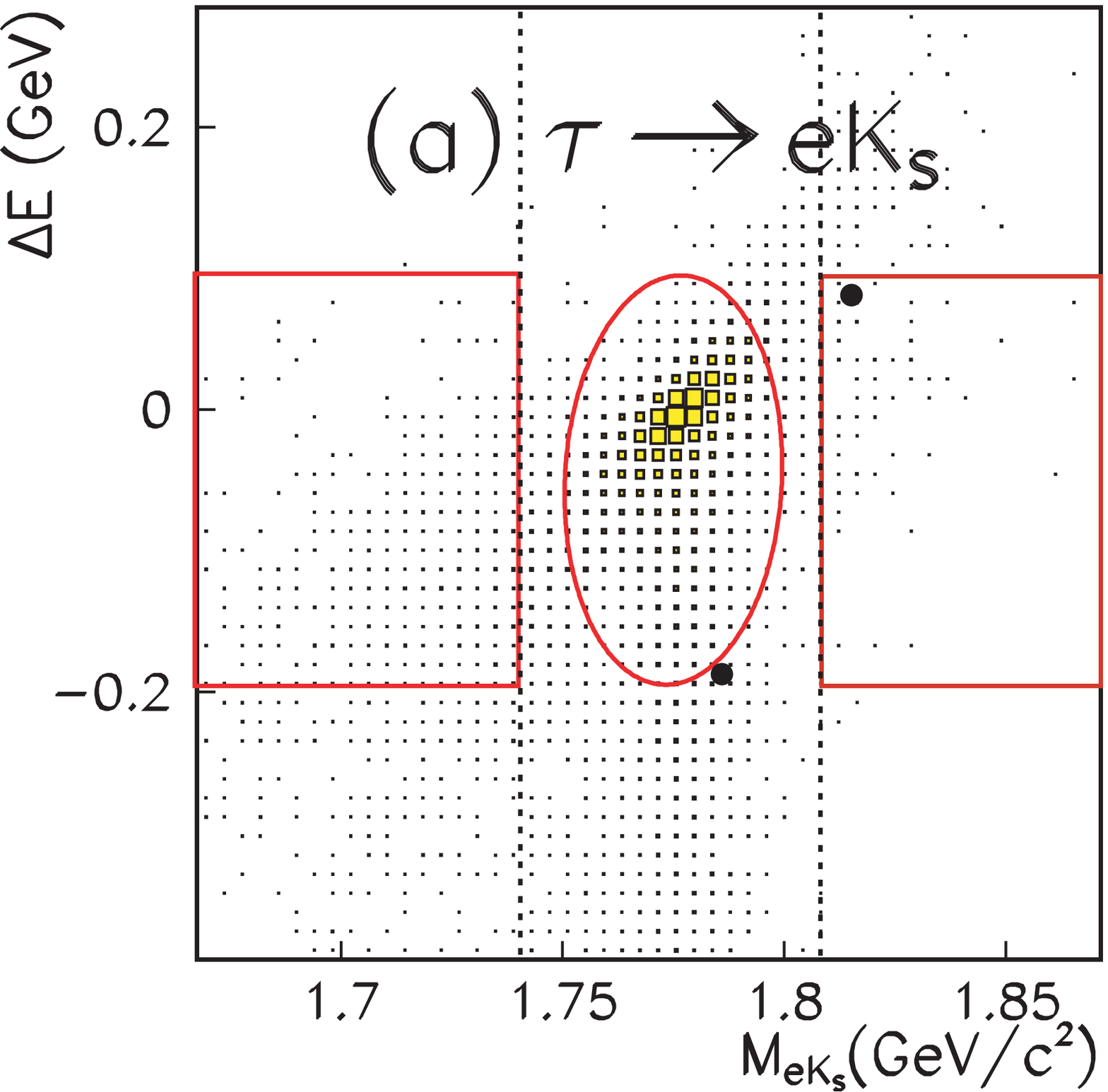}}
 \resizebox{0.496\textwidth}{0.496\textwidth}{\includegraphics
 {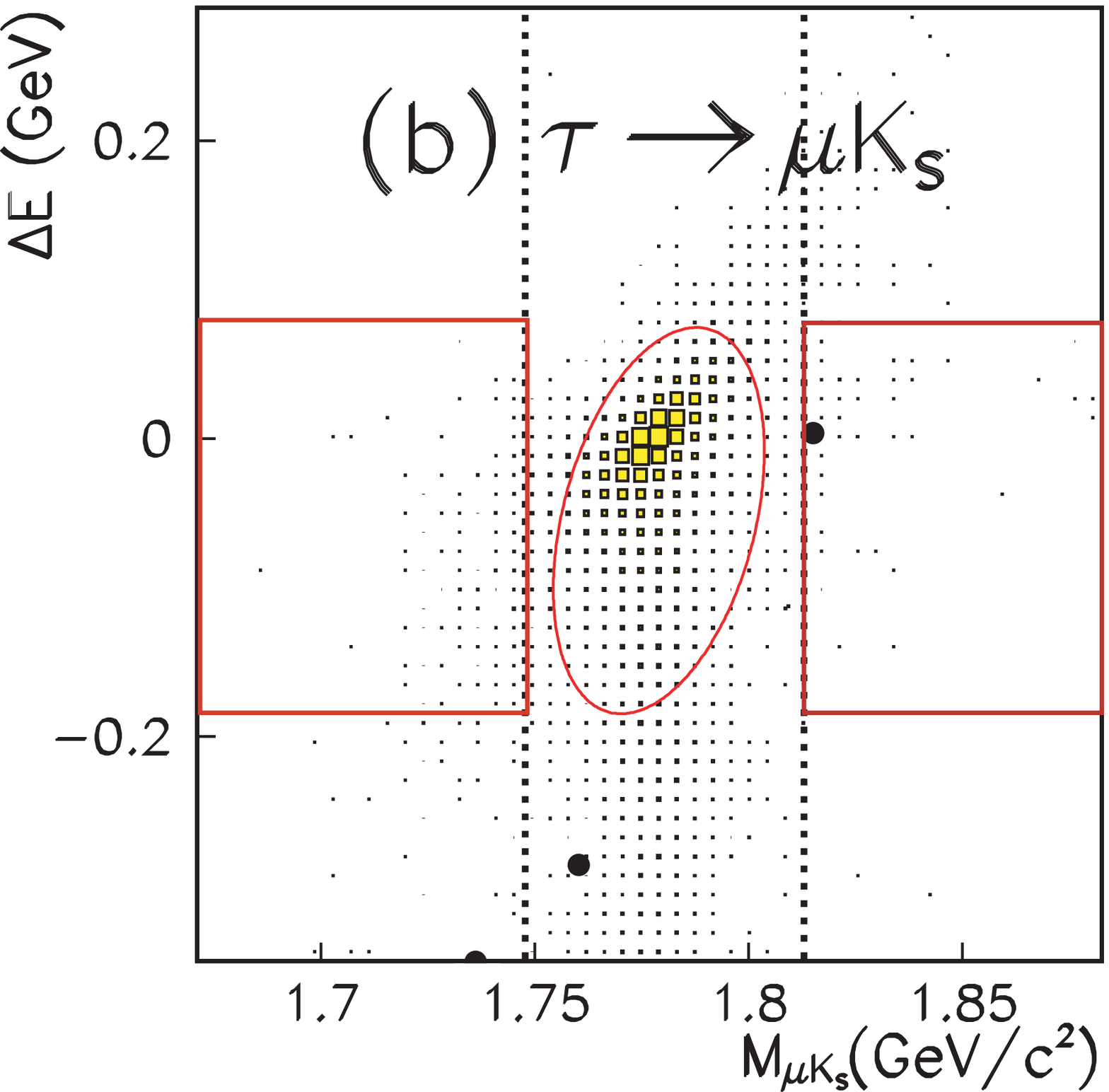}}
\caption{{{
Scatter-plots of data in the 
$M_{\ell\ks}$ -- $\Delta{E}$ plane: 
(a) and (b) correspond to
the $\pm 15 \sigma$ area for
the $\tau^-\rightarrow e^-\ks$ and $\tau^-\rightarrow \mu^-\ks$ modes, respectively.
{The data are indicated by the solid circles.}
The filled boxes show the MC signal distribution
with arbitrary normalization.
The elliptical signal region} shown by a solid curve 
is used for evaluating the signal yield.
The vertical dashed lines denote
the boundaries of
the blind regions,
while the  {rectangular areas beside the signal region} 
are
used to estimate the expected background in the elliptical region.
}}
\label{fig:3}
\end{center}
\end{figure}

{To optimize our search sensitivity, 
we select an
elliptically shaped signal region 
of minimum area
with
the same signal acceptance 
as that of a
rectangular box corresponding to
$\pm5 \sigma$ 
in the MC resolution for 
the $M_{\ell\ks}-\Delta E$ plane. 
The signal efficiencies 
after all requirements are 11.8\% for 
the $\tau^-\rightarrow e^-\ks$ 
and 13.5\% for 
the $\tau^-\rightarrow \mu^-\ks$,
respectively.}



{As there are few remaining MC background events 
{in the signal ellipse,} 
we estimate the background contribution 
{using the $M_{{\ell\ks}}$  sideband regions
defined by rectangular areas beside the signal ellipse
shown in  Fig.~\ref{fig:3}  (a) and (b).}
{Extrapolation to the signal region assumes that
the background distribution is flat in $M_{{\ell\ks}}$.}
{We find}
the expected background in the ellipse 
to be $0.2 \pm 0.2$ events for {both modes.}}
{Finally, we uncover the blinded region and find
no data events in the signal region 
of {the}
$\tau^-\rightarrow e^-\ks$ and $\tau^-\rightarrow \mu^-\ks$ modes}
(see Fig.~\ref{fig:3} (a) and (b)). 
Since no statistically significant excess of data over
the expected background in the signal region {is} observed,
we apply a frequentist approach 
to calculate upper limits {on} 
the signal {yields}~\cite{cite:FC}.
The resulting limits for 
the signal yields at 90\% confidence level, $s_{90}$,  
are 
{2.23 events} 
in both modes.
The upper limits on the branching fraction
before the inclusion of systematic uncertainties are then
calculated as
\begin{equation}
{\cal B}(\tau^- \rightarrow \ell^- \ks) 
<  \frac{s_{90}}{2 \varepsilon 
{\cal B}( \ks \rightarrow \pi^+\pi^-)
N_{\tau\tau}}
\end{equation}
where 
{${\cal B}(\ks \rightarrow \pi^+\pi^-) = 0.6895 \pm 0.0014$~\cite{PDG}} 
and 
{$N_{\tau\tau} = 251 \times 10^6$ 
is 
{the number of $\tau-$pairs
{produced} in 
281 fb${}^{-1}$ of data.
We obtain $N_{\tau\tau}$
using 
$\sigma_{\tau\tau} = 0.892 \pm 0.002$ nb, 
the $e^+e^- \rightarrow \tau^+\tau^-$ cross section  
at the $\Upsilon(4S)$ resonance 
calculated by KKMC~\cite{KKMC}.}}
The resulting values are
${\cal B}(\tau^-\rightarrow e^-\ks) < 5.5\times 10^{-8}$
and 
${\cal B}(\tau^-\rightarrow \mu^-\ks) < 4.8 \times 10^{-8}$. 

%
%

The dominant systematic uncertainties 
on 
{the detection sensitivity:} 
$2\varepsilon N_{\tau\tau}{\cal B}(\ks\rightarrow \pi^+\pi^-)$ 
{come
from $\ks$ reconstruction
and
tracking efficiencies.}
{These are 4.5\% and 4.0\%,
respectively, for both modes.}
Other sources of the systematic uncertainties
are:
the trigger efficiency (0.5\%), 
lepton identification (2.0\%),
{MC statistics (0.3\%),} 
{branching fraction of $\ks\to\pi^+\pi^-$ (0.2\%)}
and luminosity (1.4\%). 
Assuming no correlation between them,
all these uncertainties are combined in quadrature to 
give a total of {$6.5\%$}.  

While the angular distribution of $\tau^-\rightarrow \ell^-\ks$
decay is initially
assumed to be uniform {in this analysis},
it is sensitive to the lepton flavor violating interaction
structure~\cite{LFV}.
The spin correlation 
{between the $\tau$ lepton in the signal and that {in the tag side}}
must be considered.
A possible nonuniformity was taken into account by comparing
the uniform case with those assuming $V-A$ and $V+A$ interactions,
which result in the maximum possible variations.
No statistically significant difference in 
the $M_{\ell\ks}$ -- $\Delta{E}$
distribution or the efficiencies is found {compared to}
the case of the uniform distribution.
Therefore,
systematic uncertainties due to these effects 
are neglected in {the} upper limit evaluation.

{Upper limits on the branching fractions at the 90\% confidence level
including these systematic uncertainties are calculated} 
{with} the POLE program without conditioning
~\cite{cite:pole}.
The resulting upper limits on the branching fractions 
at the 90\% confidence level
are
\begin{eqnarray*}
&&{\cal B}(\tau^-\rightarrow e^-\ks) < 5.6 \times 10^{-8} \\
&&{\cal B}(\tau^-\rightarrow \mu^-\ks) < 4.9 \times 10^{-8}.
\end{eqnarray*}

\section{Discussion}
{In the $R-$parity violating SUSY scenario,
there are three kinds of terms ($\lambda$, $\lambda'$ and $\lambda''$)
with a total of 45 couplings.   
In this model,
$\tau^-$ could decay into $\ell^-\ks$  via 
tree-level scalar neutrino exchange
by {the} $\lambda\lambda'$ couplings.
Using our results,
the limits 
{on} the products $\lambda\lambda'$ 
as a function of {the} scalar neutrino mass ($M_{\tilde{\nu}}$) 
are given as~\cite{cite:rpv},} 
\begin{eqnarray*}
&&|\lambda_{i31}\lambda'_{i12}| (i=1, 2),
 |\lambda_{i31}\lambda'_{i21}| (i=2, 3)  
< 4.5 \times 10^{-4} (M_{\tilde{\nu}}/100 \mbox{GeV/$c^2$})^2
\mbox{ for }\tau^-\rightarrow e^-\ks\\
&&|\lambda_{i32}\lambda'_{i12}| (i=1, 2), 
|\lambda_{i23}\lambda'_{i21}| (i=1, 3),   
< 4.1 \times 10^{-4} (M_{\tilde{\nu}}/100 \mbox{GeV/$c^2$})^2
\mbox{ for }\tau^-\rightarrow \mu^-\ks,
\end{eqnarray*}
where $i$ is the generation number.
{These bounds are more stringent 
than
the previous {bounds} obtained
in $R-$parity violating models from $\tau^-$ decay including 
a pseudoscalar meson~\cite{cite:rpv, cite:rpv2}.}

The improved sensitivity to rare $\tau$ lepton {decays}
achieved in this work can be used to 
{constrain the new physics scale for the}
dimension-six 
fermionic effective operators involving $\tau-\mu$ flavor violation,
motivated by neutrino oscillations~\cite{cite:six_fremionic}.
{From our upper limit for 
the branching fraction of the $\tau^-\to\mu^-\ks$ decay,
lower bounds of  
36.2 TeV and  37.2  TeV
can be obtained  
for the axial-vector and 
pseudoscalar operators, respectively.}

\section{Conclusion}

{In conclusion,}
we have searched for 
{the lepton flavor {violating} decays}
$\tau^-\rightarrow\ell^-\ks$ ($\ell = e \mbox{ or } \mu$)  
using data collected 
{with} the Belle detector at the KEKB $e^+e^-$ asymmetric-energy collider.
We found no signal in either mode.
The following  upper limits on
the branching fractions 
at the 90\% confidence level
are obtained:
${\cal{B}}(\tau^-\rightarrow e^-\ks) < 5.6\times 10^{-8}$ 
and 
${\cal{B}}(\tau^-\rightarrow \mu^- \ks) < 4.9\times 10^{-8}$. 
{These results 
improve 
the search sensitivity
by factors of 16 and 19, respectively,
compared 
{to the} previous limits obtained 
{by} the {{CLEO}} experiment.}

\section*{Acknowledgments}
We thank the KEKB group for the excellent operation of the
accelerator, the KEK cryogenics group for the efficient
operation of the solenoid, and the KEK computer group and
the National Institute of Informatics for valuable computing
and Super-SINET network support. We acknowledge support from
the Ministry of Education, Culture, Sports, Science, and
Technology of Japan and the Japan Society for the Promotion
of Science; the Australian Research Council and the
Australian Department of Education, Science and Training;
the National Science Foundation of China and the Knowledge Innovation Program of Chinese Academy of Sciencies under contract No.~10575109 and IHEP-U-503; the Department of Science and Technology of
India; the BK21 program of the Ministry of Education of
Korea, and the CHEP SRC program and Basic Research program 
(grant No. R01-2005-000-10089-0) of the Korea Science and
Engineering Foundation; the Polish State Committee for
Scientific Research under contract No.~2P03B 01324; the
Ministry of Science and Technology of the Russian
Federation; the Slovenian Research Agency;  the Swiss National Science Foundation; the National Science Council and
the Ministry of Education of Taiwan; and the U.S.\
Department of Energy.

\end{document}